\documentclass[twocolumn,superscriptaddress,amssymb,amsmath,nobibnotes,aps,prd,showpacs,nofootinbib]{revtex4}
\usepackage{amsmath}
\usepackage{amsfonts}
\usepackage{amssymb}
\usepackage{graphicx}

\begin{document}
\title{Generalized energy conditions in Extended Theories of  Gravity}

\author{Salvatore Capozziello}
\email{capozzie@na.infn.it}
\affiliation{Dipartimento di Fisica, Universit\`{a} di Napoli ``Federico II'',
 Compl. Univ. di Monte S. Angelo, Edificio G, Via Cinthia, I-80126, Napoli, Italy.}
\affiliation{Istituto Nazionale di Fisica Nucleare (INFN) Sez. di Napoli, Compl. Univ. di Monte S. Angelo, Edificio G, Via Cinthia, I-80126, Napoli, Italy.}
\affiliation{Gran Sasso Science Institute (INFN), Via F. Crispi 7, I-67100, L' Aquila, Italy.}

\author{Francisco S. N. Lobo}
\email{fslobo@fc.ul.pt}
\affiliation{Centro de Astronomia
e Astrof\'{\i}sica da Universidade de Lisboa, Campo Grande, Edif\'{i}cio C8
1749-016 Lisboa, Portugal.}
\affiliation{Departamento de F\'{\i}sica, Faculdade de Ci\^encias da Universidade de Lisboa \\ 
Faculdade de Ci\^encias da Universidade de Lisboa, 
Edif\'{i}cio C8, Campo Grande, P-1749-016 Lisbon, Portugal.}
\affiliation{Instituto de Astrof\'{\i}sica e Ci\^{e}ncias do Espa\c{c}o, Universidade de Lisboa, OAL, Tapada da
Ajuda, PT1349-018 Lisboa, Portugal.}

\author{Jos\'e P. Mimoso}
\email{jpmimoso@fc.ul.pt}
\affiliation{Centro de Astronomia
e Astrof\'{\i}sica da Universidade de Lisboa, Campo Grande, Edif\'{i}cio C8
1749-016 Lisboa, Portugal.}
\affiliation{Departamento de F\'{\i}sica, Faculdade de Ci\^encias da Universidade de Lisboa \\ 
Faculdade de Ci\^encias da Universidade de Lisboa, 
Edif\'{i}cio C8, Campo Grande, P-1749-016 Lisbon, Portugal.}
\affiliation{Instituto de Astrof\'{\i}sica e Ci\^{e}ncias do Espa\c{c}o, Universidade de Lisboa, OAL, Tapada da
Ajuda, PT1349-018 Lisboa, Portugal.}

\date{\today}

\begin{abstract}

In this work, we  consider the further degrees of freedom related to curvature invariants and scalar fields in Extended Theories of Gravity (ETG). These new degrees of freedom can be recast as {\it effective fluids} that differ in nature with respect to the standard matter fluids generally adopted as sources of the field equations. It is thus somewhat misleading to apply the standard general relativistic energy conditions to this effective energy-momentum tensor, as the latter contains the matter content and a geometrical quantity, which arises from the specific ETG considered. Here, we explore this subtlety, extending on previous work, in particular, to cases with the contracted Bianchi identities with diffeomorphism invariance and to cases with generalized explicit curvature-matter couplings, which imply the non-conservation of the energy-momentum tensor. Furthermore, we apply the analysis to specific ETGs, such as scalar-tensor gravity and $f(R)$ gravity. 
Thus, in the context of ETGs, interesting results appear such as matter that may exhibit unusual thermodynamical features, for instance, gravity that retains its attractive character in the presence of large negative pressures; or alternatively, we verify that repulsive gravity may occur for standard matter.

\end{abstract}

\pacs{04.50.Kd, 98.80.-k, 95.36.+x}

\maketitle

\section{Introduction}

Modifications and extensions of General Relativity (GR) can be traced back to the early times of GR \cite{Weyl:1918,Pauli:1919,Weyl:1921,Bach:1921,Eddington:1924,Lanczos:1931}. The first extensions were aimed to unify gravity with Electromagnetism while recent interest in such modifications arises from cosmology, astrophysics and quantum gravity \cite{Schmidt:2004,OdiRev,GRGrew,report}.  In particular, cosmological observations  lead to the introduction of additional ad-hoc concepts like Dark Energy and Dark Matter, if one restricts the dynamics to the standard  Einstein theory. On the other hand, the emergence of such new ingredients of cosmic fluids could be interpreted as a first signal of a breakdown of GR on large, infrared  scales \cite{JCAP,MNRAS}. In such a way, modifications and extensions of GR become a natural alternative if such ``dark'' elements are not found out. In particular, several recent works focussed on the cosmological implications of alternative gravity since such models may lead to the explanation of the acceleration effect observed in  cosmology \cite{Capozziello:2002,Nojiri,Vollick:2003,Carroll:2005,Harko:2011nh, Harko:2012ar} and to the explanation of the missing matter puzzle observed at astrophysical scales \cite{annalen,Cap2,Borowiec:2006qr,Mar1,Boehmer:2007kx,Bohmer:2007fh,Bertolami:2007gv,
Capozziello:2013yha,Capozziello:2012}.

While it is very natural to extend Einstein's gravity to theories with additional geometric degrees of freedom \cite{Hehl:1976,Hehl:1995,Trautman:2006}, recent attempts focussed on the idea of modifying the gravitational Lagrangian  leading to higher-order field equations.  Due to the increased complexity of the field equations, a huge amount of works considered  some formally equivalent theories, in which a reduction of the order of the field equations can be  achieved by considering the metric and the connection as independent objects \cite{francaviglia,olmo}. 
However, a  concern which arises with generic extended and modified gravity theories is linked to the initial value problem and the definition of the energy conditions. It is unclear if standard methods can be used in order to tackle these problems in any theory. Hence it is doubtful that the full Cauchy problem can be properly addressed, if one takes into account the results already obtained in GR. 
On the other hand, being alternative gravities, like GR, gauge theories, the initial value formulation and the energy conditions  depend on suitable constraints and  gauge choices  \cite{Teyssandier:Tourrenc:1983,Noakes}.
A different approach is possible  showing  that the Cauchy problem for alternative gravities can be  well-formulated and well-posed in vacuo, while it can be, at least, well-formulated for various form of matter fields like perfect-fluids, Klein-Gordon and  Yang-Mills fields \cite{vignolo}. A similar situation also holds for the energy conditions which can strictly depend on the kind of fluids adopted as sources in the field equations. 

In fact, there are serious problems of deep and fundamental principle at the semi-classical level and certain classical systems exhibit seriously pathological behaviour, in particular, the classical energy conditions are typically violated by semiclassical quantum effects \cite{Barcelo:2002bv}. 
In this context, some effort has gone into finding possible semiclassical replacements for the classical energy conditions \cite{Ford:1990id}. Recently, classical and quantum versions of a ``flux energy condition'' (FEC and QFEC) were developed based on the notion of constraining the possible fluxes measured by timelike observers \cite{Martin-Moruno:2013sfa}. It was shown that the naive classical FEC was satisfied in some situations, and even for some quantum vacuum states, while its quantum analogue (the QFEC) was satisfied under a rather wide range of conditions. Furthermore, several nonlinear energy conditions suitable for use in the semiclassical regime were developed, and  it was shown that these nonlinear energy conditions behave much better than the classical linear energy conditions in the presence of semiclassical quantum effects \cite{Martin-Moruno:2013wfa}.

However, in the context of alternative theories of gravity that in a wide sense {\it extend} GR, the issue of the energy conditions is extremely delicate. Note that the further degrees of freedom carried by these Extended Theories of Gravity (ETGs) can be recast as generalized 
{\it effective fluids}
that differ in nature with respect to the standard matter fluids generally adopted as sources of the field equations \cite{report}. This approach has been extensively explored in the literature, namely, the energy conditions have been used to constrain $f(R)$ theories of gravity \cite{PerezBergliaffa:2006ni, Santos:2007bs, Atazadeh:2008mh} and extensions involving nonminimal curvature-matter couplings \cite{Bertolami:2009cd, Wang:2010zzr,Garcia:2010xb, MontelongoGarcia:2010xd, Wang:2012rw, Wang:2012mws, Wu:2014yya}; bounds on modified Gauss-Bonnet $f(G)$ gravity from the energy conditions have also been analysed \cite{Wu:2010zzm, Garcia:2010xz, MontelongoGarcia:2010ip}, and with a nonminimal coupling to matter \cite{Banijamali:2011up}; the recently proposed $f(R,T)$ gravity models \cite{Harko:2011kv}, where $T$ is the trace of the energy-momentum tensor and $R$ is the curvature scalar, have also been tested using the energy conditions \cite{Alvarenga:2012bt, Sharif:2012gz, Sharif:2012ce}; and constraints have also been placed \cite{Sharif:2013kga} on the $f(R, T, R_{\mu\nu} T^{\mu\nu})$ extension \cite{Haghani:2013oma, Odintsov:2013iba}; bounds have been placed on modified teleparralel gravity \cite{Liu:2012fk, Bohmer:2011si, Jamil:2012ck}; and the null energy condition violations have been studied in bimetric gravity \cite{Baccetti:2012re}.

However, one should add a cautionary note of the results obtained in the literature, such as the majority of those considered above have recast the further degrees of freedom carried by these ETGs as generalized {\it effective fluids} that differ in nature with respect to the standard matter fluids generally adopted as sources of the field equations \cite{report}.
Note that while standard fluids (e.g., perfect matter fluids), generally obey standard equations of state (and then one can define every thermodynamic quantity such as the adiabatic index, temperature, etc), these ``fictitious'' fluids can be related to scalar fields or further gravitational degrees of freedom. In these cases, the physical properties can result ill-defined and the energy conditions could  rigorously work as in GR. The consequences of such a situation can be dramatic since the causal and geodesic structures of the theory could present  serious shortcomings as well as the energy-momentum tensor could not be consistent with the Bianchi identities and the conservation laws.

This paper is outlined in the following manner. In Section \ref{secII}, we briefly review the energy conditions in GR and discuss the geometrical implications of such conditions. Section \ref{secIII} is devoted to set the energy conditions in ETGs by considering, in particular,  the contracted Bianchi identities, the non-conservation of the energy-momentum tensor, the propagation equations and  the role of conformal transformations. In Section \ref{secIV}, we take into account some particular theories, i.e., scalar-tensor theories and $f(R)$ gravity, where $R$ is the Ricci scalar. Finally, we discuss our results and draw some conclusions in Sec. \ref{sec:concl}.

\section{The energy conditions in General Relativity}
\label{secII}

In GR, the Einstein field equation govern the interplay between the geometry of the spacetime and the matter content. More specifically, the field equation is given by
\begin{equation}
G_{ab} = 8\pi G\, T_{ab} \, ,
\end{equation}
where the energy-momentum tensor of the matter fields, $T_{ab}$ is related to the Einstein tensor $G_{ab}\equiv R_{ab}-\frac{1}{2}\,g_{ab}R$, with $R_{ab}$ the Ricci tensor, which is defined as the trace of the Riemann curvature tensor ${R^d}{}_{adb}=R_{ab}$, and $R={R^a}_a$. Thus, the imposition of specific conditions on $T_{ab}$ are translated into corresponding conditions on the Einstein tensor $G_{ab}$. 
Note that the Einstein equations can also be cast as conditions on the Ricci tensor, that is 
 \begin{equation}
 R_{ab}= 8\pi G\, \left( T_{ab}-\frac{1}{2}\,Tg_{ab}\right)\,.
 \end{equation}
In this form, the role of energy-matter is more relevant.

In general, in considering the energy conditions, we take into account a congruence of timelike curves whose tangent 4-vector is, for instance, $W^a$. The latter represents the velocity vector of a family of observers. One may also consider a field of null vectors, $k^a$, so that $g_{ab}\, k^ak^b=0$ implies that $G_{ab}k^a k^b =R_ {ab}k^ak^b$. 

These choices  enable us to identify the physical quantities measured by the observers related to the timelike vector $W^a$. Indeed, with respect to the latter vector field $W^a$, the energy-momentum tensor can then be decomposed as
\begin{equation}
T^{ab} = \rho\, W^a W^b + p \, (g^{ab}+W^aW^b) + \Pi^{ab} + 2q^{(a} W^{b)} \; , \label{EMT_decomposition}
\end{equation}
where  $\rho$ and $p$ are the energy-density and the (isotropic) pressure measured by the observers moving with velocity $W^a$, $\Pi^{ab}$ is the anisotropic stress tensor, and $q^a$ is the current vector of the heat/energy flow. These quantities are given by the following relations 
\begin{eqnarray}
\rho &=& T_{cd}\,W^cW^d \,, \\
3p&=&T_{cd}\,h^{cd} \,, \\
\Pi^{ab} &=& \left(h^{ac}h^{bd}-\frac{1}{3}h^{ab}h^{cd}\right)T_{cd} \,, 
  \label{defPI} \\
q^a &=& W^c T_{cd}h^{ad} \,,
\end{eqnarray}
respectively, where $h^{ab} = g^{ab}+W^aW^b$ is the metric induced on the spatial hypersurfaces orthogonal to $W^a$. Throughout this work, we adopt the $(-+++)$ signature convention and the speed of light is $c=1$.

\subsection{The classification of energy conditions}

The energy conditions are defined by considering contractions of timelike and null vectors with respect to the Ricci, Einstein and energy-momentum tensors \cite{Hawking:1973uf}. They can be classified as follows.
  
\begin{itemize}

\item The {\em weak energy condition} is defined as
\begin{equation}
T_{ab}\, W^a W^b  \ge 0\; , \label{en_cond_weak}
\end{equation}
where $W^a$ is a timelike vector, i.e., $ W^a W_a =-1$. From Eq. (\ref{EMT_decomposition}), we verify that this entails that $\rho \ge 0$. As presented by Hawking and Ellis \cite{Hawking:1973uf}, such a condition is equivalent to establishing that the energy density measured by any observer is non-negative. It is straightforward to demonstrate that any standard matter  fluid is consistent with such a condition.
Through the Einstein field equations where the curvature of space-time is considered, condition (\ref{en_cond_weak}) translates into 
\begin{equation}
G_{ab}\, W^a W^b  \ge 0
\end{equation}
which is equivalent to 
\begin{equation}
R_{ab} \, W^a W^b \ge - \frac{R}{2} \; .
\end{equation}
and also to
\begin{equation}
R_{ab} \, W^a W^b \ge -4\pi G\, (\rho-3p) \; .
\end{equation}
Here we have used the fact that, from Eq. (\ref{EMT_decomposition}), we can recast the Einstein equations as
\begin{eqnarray}
R^{ab} = 8\pi G\, \Big[ \frac{\rho+3p}{2}\, W^a W^b 
+ \Pi^{ab} + 2q^{(a} W^{b)}
\nonumber  \\
+ \frac{\rho-p}{2}\, (g^{ab}+W^aW^b) 
 \Big] \;. \label{EFE_decomposition2}
\end{eqnarray}

\item The {\em dominant energy condition} states that, in addition to the condition (\ref{en_cond_weak}), one also has that $T^{ab}W_b$ is a non-spacelike vector, where as before $ W^a$ is a timelike vector, so that $W^a W_a =-1$.  This corresponds to having a local energy flow vector which is non-spacelike in addition to the non-negativity of the energy density. In this sense, the causal structure of the space-time is determined.

\item The {\em null energy condition}  states that
\begin{equation}
T_{ab}\, k^a k^b  \ge 0\; , \label{en_cond_null}
\end{equation}
where $k^a$ is a null vector, i.e., $ k^a k_a =0$. This implies $R_{ab}k^a k^b\ge 0$, through the Einstein field equation. 
A very useful meaning of this condition is that it's violation implies that
the Hamiltonian of the corresponding system is necessarily unbounded
from below (we refer the reader to \cite{Sawicki:2012pz} for more details).

\item The {\em strong energy condition} is given by
\begin{equation}
T_{ab}\, W^a W^b  \ge \frac{1}{2}T\,W^aW_a\; , \label{en_cond_strong}
\end{equation}
where $W^a$ is a timelike vector. Alternatively, in GR and through the Einstein field equations, the above inequality takes the form 
\begin{equation}
R_ {ab} \, W^a W^b \ge 0
\end{equation}
which, as we will see in what follows through the Raychaudhuri equation, states that gravity must be attractive. 

\end{itemize}
Summarizing, such conditions define the causal structure, the geodesic structure and the nature of the gravitational field in a space-time filled by a standard fluid matter endowed with a regular equation of state.

\subsection{Geometrical implications of the energy conditions }

The geometrical implications of the energy conditions can be put in evidence  as soon as  we consider the decomposition \cite{Ellis}
\begin{equation}
\nabla_b W_a = \sigma_{ab} +\frac{\theta}{3}h_ {ab}+\omega_{ab} -\dot{u}_a W_b \,,
\label{3+1_decomp}
\end{equation}
with the following definitions
\begin{eqnarray}
h_{ab} &=& g_{ab}+W_a\,W_b  \,, \\
\sigma_{ab} &=& h_{(a}^c\nabla_c W_d h_{b)}^d - \frac{h_{ab}}{3}\,h_a^c\nabla_c W_d h^{ad} \,,  \\ 
\theta &=& h_a^c\nabla_c W_d h^{ad}  \,, \\
\omega_ {ab} &=&  h_{[a}^c\nabla_c W_d h_{b]}^d  \,, \\
\dot W^a &=& W^b \nabla_b W^a \; ,
\end{eqnarray}
respectively, where we have considered all possible combinations of the metric tensor and timelike vectors. Here $h_{ab}$ is the {\it projection tensor}, $\sigma_{ab}$ is the {\it shear tensor}, $\theta$ is the {\it expansion scalar}, $\omega_{ab}$ is the vorticity tensor.  Note that $h_{ab}$ is orthogonal to $W^a$, $W^a\,h_{ab} =0$, and hence it is the metric induced on the 3-hypersurfaces orthogonal to $W^a$, as mentioned before.

Equipped with the latter kinematical quantities whose contractions give rise to the so-called {\it optical scalars} \cite{falco}, we derive, from the Ricci identities\footnote{The Ricci identities prescribe that $\nabla_c\nabla_d u^a-\nabla_d\nabla_c u^a = {R^a}_{bcd}\,u^b$, for any vector field $u^a$.}, the following relations
\begin{equation}
\dot \theta + \frac{\theta^2}{3}+2\,(\sigma^2-\omega^2) - \dot{u}^a{}_{;a}=- R_ {ab} \, W^a W^b  \;, \label{Ray_1}
\end{equation}
and
\begin{eqnarray}
{h_a}^f \,{h_b}^g\,\left[(\sigma_{fg})\dot{}\, - \dot{W}_{(f;g)}\right] =  \dot{W}_a\,  \dot{W}_b - \omega_a\,\omega_b -\sigma_{af}{\sigma^f}_b 
 \nonumber  \\
  - \frac{2}{3}\,\theta\,\sigma_{ab} 
-h_{ab}\left(-\frac{1}{3}\,\omega^2 - \frac{2}{3}\,\sigma^2 - \frac{1}{3}\,{\dot{W}^c}_c\right) 
\nonumber  \\ 
 + \frac{1}{2}\left(h^{ac}h^{bd}-\frac{1}{3}h^{ab}h^{cd}\right)\, \left( R_ {cd}-\frac{1}{2}g_{cd}R\right) \label{Shear_propag_1}
 \end{eqnarray}
 \begin{eqnarray}
{h^a}_b\left[ \exp{\left(\frac{2}{3}\int\,\theta\,dt \right)}\,\omega^b\right]{\dot{}} = {\sigma^a}_b \left[ \exp{\left(\frac{2}{3}\int\,\theta\,dt \right)}\,\omega^b\right] 
\nonumber \\ 
+\frac{1}{2}\left[ \exp{\left(\frac{2}{3}\int\,\theta\,dt \right)} \right]\,\eta^{abcd}\,W_b\,\dot{W}_{(c;d)} \; , \label{Vortice_propag_1}
\end{eqnarray}
where Eq. (\ref{Ray_1}) is the so-called {\it Raychaudhuri Equation}.
It is important to emphasize that Eqs. (\ref{Ray_1})--(\ref{Vortice_propag_1}) only carry a geometrical meaning, as they are directly derived from the Ricci identities. It is only when we choose a particular theory that we establish a relation between quantities that appear in their right-hand sides, such as $R_ {ab} \, W^a W^b $ in Eq. (\ref{Ray_1}), and the energy-momentum tensor describing matter fields.

For instance, let us consider a null congruence $k^a$ and a vanishing  vorticity $\omega_{ab}=0$.  The Raychaudhuri Eq. (\ref{Ray_1}) reduces to
\begin{equation}
\frac{{\rm d} \theta}{{\rm d}v} = - \left[\frac{\theta^2}{3}+2\sigma^2 +R_{ab}k^ak^b\right]\; ,
\end{equation}
where $v$ is an affine parameter along the null geodesics. This means that, in GR, it is possible to associate  the null energy condition with the focusing (attracting) characteristic of the spacetime geometry. Gravitational lensing is a very important application of this feature as widely discussed in \cite{falco}. 

\section{The problem of energy conditions in Extended  Theories of Gravity}
\label{secIII}

In the context of ETGs, consider the following generalized gravitational field equations, which encapsulates a large class of interesting cases
\begin{equation}
g_1(\Psi^i)\, \left( G_{ab} +H_ {ab}\right)= 8\pi G\, g_2(\Psi^j) \, T_{ab} \, , \label{ETG_EFEcoup}
\end{equation}
where the factors $g_1(\Psi^i)$ modifies the coupling with the matter fields in $T^{ab}$ and $g_2(\Psi^i)$ incorporates explicit curvature-matter couplings of the gravitational theory considered \cite{Harko:2012ar,Bertolami:2007gv}; $\Psi^j$ generically represents either curvature invariants or other gravitational fields, such as scalar fields, contributing to the dynamics of the theory. The additional tensor $H_{ab}$ represents an additional geometric term with regard to GR that encapsulates the geometrical modifications introduced by the extended theory under consideration. 

Note, that GR is immediately recovered by imposing $H_{ab}=0$, $g_1(\Psi^i)=g_2(\Psi^i)=1$. In this sense we are dealing with Extended Theories of Gravity, in that the underlying hypothesis is that GR (and its positive results) can be recovered as a particular case in any ``extended'' theory of gravitation~\cite{ETG_2_GR}. 

\subsection{Contracted Bianchi identities and diffeomorphism invariance}

Consider the specific case of $g_1(\Psi^i)=g(\Psi^i)$ and $g_2(\Psi^i)=1$, so that the field equation (\ref{ETG_EFEcoup}) reduces to
\begin{equation}
g(\Psi^i)\, \left( G_{ab} +H_ {ab}\right)= 8\pi G\, T_{ab} \; .\label{ETG_EFE1}
\end{equation}

Taking into account the contracted Bianchi identities and the diffeomorphism invariance of the matter action,  which implies the covariant conservation of the energy-momentum tensor, $\nabla_b T^{ab}=0$, one deduces the following conservation law
\begin{equation}
\nabla_b H^{ab} = - \frac{8\pi G}{g^2} T^{ab}\, \nabla_b g\; .
\end{equation}
Note that from Eq. (\ref{ETG_EFE1}) in order to have an extended Bianchi identity $\nabla_b H^{ab} =0$, for a non diverging value of the coupling $g$, we must have vacuum and therefore $G_{ab}=-H_{ab}$.

Now, an imposition of specific energy conditions on the energy-momentum tensor $T^{ab}$ carries over the conditions to the combination of $G_ {ab}$ and $H_ {ab}$ and not just for the Einstein tensor. Thus, in the context of ETGs, it is not possible to obtain a simple geometrical implication from the conditions imposed. For instance, in GR, suppose that the strong energy condition holds. This would mean that $R_ {ab} \, W^a W^b  \ge 0$, and consequently through the Einstein field equation we would have $\rho+3P\ge 0$. On the other hand, this entails gravity with an attractive character, since given Eq. (\ref{Ray_1}), one verifies that the geodesics are focusing \cite{Hawking:1973uf}. 
However, in the ETG case under consideration, this condition just states that 
\begin{equation}
g(\Psi^i)\,(R_{ab}+H_ {ab}-\frac{1}{2}g_{ab} H)\,W^aW^b \ge 0\; , \label{en_cond_strong_2}
\end{equation}
which does not necessarily entail $R_ {ab} \, W^a W^b  \ge 0$, so that one cannot conclude that the attractive nature of gravity is equivalent to the satisfaction of the strong energy condition, in the particular ETG under consideration \cite{Capozziello:2013vna}.

However, in the literature, it is common practise to transport the term $H^{ab}$ to the right-hand-side of the gravitational field equation, and write the latter as a modified Einstein field equation, namely,
\begin{equation}
 G_{ab}= 8\pi G \;T_{ab}^{\rm eff} \;, \label{mod_EFE}
\end{equation}
where $T_{ab}^{\rm eff}$ is considered as an effective energy-momentum tensor, defined by $T_{ab}^{\rm eff}=T_{ab}/g-8\pi G \, H_{ab}$. Thus, the meaning which  is attributed to the energy conditions is the satisfaction of some inequality by the combined quantity $T^{ab}/g-H^{ab}$.  It is therefore somewhat misleading to call these impositions as energy conditions since they do not emerge only from $T^{ab}$ but from a combined quantity where we are dealing with a geometrical $H^{ab}$ as an additional stress-energy tensor. 
Indeed, we emphasize that $H^{ab}$ is a geometrical quantity, in the sense that it can be given by geometrical invariants as $R$ or scalar fields different from ordinary matter fields.

However, if the ETG under consideration allows an equivalent description  upon an appropriate conformal transformation, it then becomes justified to associate the transformed $H^{ab}$ to the redefined $T^{ab}$ in the conformally transformed Einstein frame. This is, for instance, the case for  scalar-tensor gravity theories, and for instance in $f(R)$ gravity \cite{report}. Indeed, conformal transformations play an extremely relevant role in the discussion of the energy conditions. In particular, they allow to put in evidence the further degrees of freedom coming from ETGs under the form of curvature invariants and scalar fields. More specifically, several generalized  theories of gravity  can be redefined as GR plus a number of appropriate fields coupled to matter by means of a conformal transformation in the so-called Einstein frame. 

In fact, in scalar-tensor gravity, in the so-called Jordan frame one has a separation between the geometrical terms and the standard matter terms that can be cast as in Eq. (\ref{ETG_EFE1}), where $H_{ab}$  involves a mixture of both the scalar and tensor gravitational fields.
A main role in this analysis is played by recasting the theory, by conformal transformations,  in the Einstein frame where matter and geometrical quantities can be formally dealt exactly such as in GR.  However, the energy conditions can assume a completely different meaning going back to the Jordan frame and then they could play a crucial role in identifying the physical frame as firstly pointed out in \cite{magnano}. 
Although, it is completely clear that different ``frames'' just correspond to field redefinitions all of which are equally physical.

Now, under a suitable conformal transformation the field equations can be recast as 
\begin{equation}
\tilde{G}_ {ab}= \tilde{T}^M_{ab}+\tilde{T}^\varphi_{ab} \,, \label{STT_EF}
\end{equation}
where $\tilde{T}^M_{ab}$ is the transformed energy-momentum of matter, and $\tilde{T}^\varphi_{ab}$ is an energy-momentum tensor for the redefined scalar field $\varphi$ which is coupled to the matter. It thus makes sense to consider the whole right-hand side of (\ref{STT_EF}) as an effective energy-momentum tensor.
Then one finds results where one draws conclusions about the properties of $G_ {ab}$ such whether it focuses geodesics directly from those conditions holding on $T_{ab}^{\rm eff}$, where $T_{ab}^{\rm eff}=\tilde{T}^M_{ab}+\tilde{T}^\varphi_{ab}$. This ignores the fact that $H_{ab}$ originally possesses a geometrical character, and thus the conclusions may be too hasty if not supported by the physical analysis of sources. We refer the reader to \cite{Capozziello:2013vna} for a detailed analysis on this issue.

\subsection{Non-conservation of the energy-momentum tensor}

A main role in the formulation of the correct energy conditions for  ETGs is played by the contracted  Bianchi identities that guarantee specific conservation laws. In fact, being  $\nabla_b G^{ab}=0$, the physical features of $H^{ab}$ can be derived. On the other hand, the Bianchi identities guarantee the self-consistency of the theory. However, an interesting class of extended theories of gravity that exhibit an explicit curvature-matter coupling have recently been proposed in the literature \cite{Harko:2012ar,Bertolami:2007gv}. The latter coupling imply a general non-conservation of the energy-momentum tensor, and consequently a trademark of these specific ETGs is non-geodesic motion \cite{Harko:2012ar,Bertolami:2007gv}. 

We will briefly analyse these theories in the formalism outlined above. In order to incorporate the explicit curvature-matter coupling, consider the field equation given by Eq. (\ref{ETG_EFEcoup}). Note that in ETGs of the form (\ref{ETG_EFEcoup}) in the presence of the non-conservation of the 
energy-momentum tensor, the contracted Bianchi identities yield
\begin{equation}
\nabla_b H^{ab} = \nabla_b \left(\frac{T^{ab}}{\bar{g}}\right) \,,
  \label{conserv_coupl}
\end{equation}
where the factor $\bar{g}=g_1/g_2$ is defined, and we have considered that $8\pi G=1$ for notational simplicity.
Now, Eq. (\ref{conserv_coupl}) implies the following relationship
\begin{eqnarray}
\nabla_b T^{ab} &=& \bar{g}\,\nabla_b H^{ab}+\left(\frac{\nabla_b \bar{g}}{\bar{g}}\right) T^{ab}  \nonumber \\ 
&=& \nabla_b( \bar{g}\,H^{ab}) +\left(\frac{\nabla_b \bar{g}}{\bar{g}}\right) \,\left[ T^{ab} -(\bar{g}\,H^{ab}) \right].
\end{eqnarray}
Thus a trademark of these specific class of ETGs is that the matter fields do not, in general, follow the geodesics of space-time \cite{Harko:2012ve}.

Let us we introduce the following useful definitions 
\begin{eqnarray}
\tilde\rho &=& (\bar{g}\,H_{cd})\,W^cW^d \,, \label{redef_tilderho} \\ 
3\tilde p&=&(\bar{g}\,H_{cd})\,h^{cd}   \,, \label{redef_tilderho-b} \\
\tilde\Pi^{ab} &=& \left(h^{ac}h^{bd}-\frac{1}{3}h^{ab}h^{cd}\right)(\bar{g}\,H_{cd}) \,, \label{redef_tilderho-c} \\
\tilde q^a &=& W^c \,(\bar{g}\,H_{cd})\,h^{ad} \,. \label{redef_tilderho-d}
\end{eqnarray} 
We derive
\begin{equation}
\dot \rho + (\rho+p) \nabla_b W^b + W_a \,\nabla_b \Pi^{ab}= W_a \,\nabla_b \bar{g}\,H^{ab} \,,
\end{equation}
so  that the departure from the usual conservation equations depends on the term
\begin{equation}
W_a \,\nabla_b \left(\bar{g} H^{ab}-\Pi^{ab}\right) \; .
\end{equation}
Therefore in what regards this balance equation, the term $\bar{g} H_ {ab}$ plays a role which is analogous to that of the anistropic stress tensor $\Pi_{ab}$, given by Eq. (\ref{defPI}). We can recast the latter equations as
\begin{eqnarray}
\dot \rho + (\rho+p) \theta + \Pi^{ab}\,  \sigma^{ab}+\nabla_b q^b + \dot W_a\,q^a =
  \nonumber \\ 
= \left[\dot {\tilde\rho} + (\tilde\rho+\tilde p) \theta
+ \tilde\Pi^{ab}\,  \sigma^{ab}+\nabla_b \tilde q^b + \dot W_a\,\tilde q^a \right] 
  \nonumber \\  
 + \left(\frac{\dot{\bar{g}}}{\bar{g}}\right) \,\left( \tilde\rho-\rho \right)  +\left(\frac{\nabla_b \bar{g}}{\bar{g}}\right) \,\left( \tilde q^b -q^b\right)  ,
\end{eqnarray}
or as
\begin{eqnarray}
\dot \rho- \dot {\tilde\rho} + \left[(\rho-\tilde\rho )+(p-\tilde p)\right] \theta = - 
\left(\Pi^{ab}-\tilde\Pi^{ab} \right)\,  \sigma^{ab} 
    \nonumber  \\
- \nabla_b \left(q^b -{\tilde q}^b \right)
-\dot W_a\,\left(q^a-{\tilde q}^b\right) 
+ \left(\frac{\dot{\bar{g}}}{\bar{g}}\right) \,\left( \tilde\rho-\rho \right)
 \nonumber  \\ 
  +\left(\frac{\nabla_b \bar{g}}{\bar{g}}\right) \,\left( \tilde q^b -q^b\right) \,.
\end{eqnarray}

Analogously, we derive an equation for the acceleration $\dot W^a$. We obtain the following relationships
\begin{eqnarray}
&&\left[(\rho-\tilde\rho)+(p-\tilde p)\right] \,\dot W^a +h_a^b 
\left[\nabla_b(p-\tilde p)\right] 
   \nonumber \\
&& = - 
h_a^c\nabla_b(\Pi^b_c-\tilde{\Pi}^b_c) - h_a^c\, (\dot q_c 
-\dot{\tilde{q}}_c)    + \left(\frac{\nabla_b  \bar{g}}{\bar{g}}\right)  \times
    \nonumber \\ 
&& \times  \left\{ \left( p-\tilde p 
\right) h_a^b + \,\left[\left(\Pi_a^b -\tilde\Pi_a^b \right)-\left(\tilde q_a -
q_a\right)W^b\right] \right\}.
\end{eqnarray}
These equations show how the $\bar{g}H^{ab}$ term modifies the standard energy density conservation equation and the generalized Navier-Stokes equation for the acceleration, both derived  from the contracted Bianchi identies. It is important to emphasize that,  although the contracted Bianchi  identities are  geometrical relations in their essence, and hence do not depend on the specific gravitational theory under consideration, when we translate them into equations governing the behavior of the matter fields,  the choice of the theory intervenes. This happens in association with the  $\bar{g}H^{ab}$ terms, that is with the tilded quantities that we have defined in the Einstein frame.  In summary, the validity of the contracted Bianchi identities selects suitable theories and may allow the definition of self-consistent energy conditions.

\subsection{Propagation equations and Extended Theories of Gravity}

In the present subsection, we consider the specific case of $\bar{g}=g$, i.e., $g_2=1$, and consequently the covariant conservation of the energy-momentum tensor.
The role of propagation equations deserve a particular discussion in this context.
We have already written the propagation equations for the expansion $\theta$, for the shear $\sigma_{ab}$ and for the vorticity  $\omega_{ab}$, that is  Eqs. ~(\ref{Ray_1})-(\ref{Vortice_propag_1}), and have pointed out that these equations do not reflect the particular gravitational theory under consideration since they are derived directly from the 3+1 decomposition of the Ricci identities that come from the Riemann tensor. 

The prescription for a given gravitational theory enters into play when we replace quantities such as $R_ {ab} \, W^a W^b $
into the Raychaudhuri Eq. (\ref{Ray_1}).  For the theories under consideration here, the latter geometrical quantity is replaced by the inequality (\ref{en_cond_strong_2}), which, according to the  definition (\ref{redef_tilderho}) (recall that in the present context we have $\bar{g}=g$, i.e., $g_2=1$, and the covariant conservation of the energy-momentum tensor), only involves the energy density of matter and that given by the latter equation. However, when we consider the shear propagation equation, the role of the particular  ETG comes out by  replacing  $\frac{1}{2}\left(h^{ac}h^{bd}-\frac{1}{3}h^{ab}h^{cd}\right)\, \left( R_ {cd}-\frac{1}{2}g_{cd}R\right)$. We thus have
\begin{eqnarray}
\frac{1}{2}\left(h^{ac}h^{bd}-\frac{1}{3}h^{ab}h^{cd}\right)\, \left( R_ {cd}-\frac{1}{2}g_{cd}R\right) =
 \nonumber \\
= \frac{1}{2}\left(h^{ac}h^{bd}-\frac{1}{3}h^{ab}h^{cd}\right)\,\left(-H_{ab}+\frac{T_{ab}}{g}\right) 
    \nonumber \\
= \frac{1}{g}\, \left(-\tilde\Pi^{ab}+ \Pi^{ab}\right) \; .
\end{eqnarray}

In general, the discussion of the energy conditions in ETGs is made in relation to the spatially homogeneous and isotropic FLRW universes, which implies that $\sigma_{ab}=0$ and $\omega_{ab}=0$\footnote{One also has the vanishing of the electric and magnetic parts of the Weyl tensor $C_{abcd} $, $E_{ab}=C_{acbd}\,W^cW^d$ and $H^\ast_{ab}=\frac{1}{2}\,{\eta_{ac}}^{gh}\,C_{ghbd}\,W^cW^d$, respectively, where ${\eta}^{abcd}$ is the totally-skew symmetric pseudotensor.}. One question which is then of interest is to assess the possible role of the ETG theories in perturbing the universe away from its Friedmann state. Clearly this depends on the term $\tilde\Pi^{ab}$ being non-vanishing.
The interesting result that we want to put forward is that in theories like $f(R)$ gravity  and scalar-tensor gravity, the quantity  
$\tilde\Pi^{ab}$ is vanishing and so they do not introduce any modification with respect  to GR in the shear propagation equation. If the shear starts vanishing, it remains so. Indeed, theories where $\tilde\Pi^{ab}\neq 0$ exist (e.g. inhomogeneous cosmologies \cite{krasinski}) but we do not consider them in the present context.

\section{Examples of Extended Theories of Gravity}\label{secIV}

Taking into account the above discussion, the correct identification of the function $g_i(\Psi^j)$ ($i=1,2)$, and the tensor $H_{ab}$ defined in Sec. \ref{secIII} enables one to formulate the energy conditions for any ETG. Recall that the functions $g_i(\Psi^j)$ are related to the gravitational coupling that can be non-minimal, and the tensor $H_{ab}$ is the contribution to the effective energy-momentum tensor containing the further degrees of freedom of the ETG. Below, we give some specific examples of theories that fit well in the context of the above discussion.

\subsection{Scalar-Tensor gravity}

In this subsection, we extend and complement the analysis outlined in \cite{Capozziello:2013vna}. The scalar-tensor gravity~\cite{ST}, to which Brans-Dicke is the archetype, can be based on the action
\begin{equation}
S=\frac{1}{16\pi} \int \sqrt{-g} d^4x\, \left[\phi R - \frac{\omega(\phi)}{\phi}
\phi_{,\mu} \phi^{,\mu} + 2 \phi \lambda(\phi)\right] + S_M \,,
\end{equation}
where the gravitational coupling is assumed variable and a self-interaction potential is present; $S_M$ is the standard matter part. Varying this action with respect to the metric $g_{ab}$  and the scalar field $\phi$ yields the  field equations
\begin{eqnarray}
R_{ab}-\frac{1}{2}g_{ab}\, R -\lambda(\phi)\, 
g_{ab}=\frac{\omega(\phi)}{\phi^2}\;
\left[\phi_{;a}\phi_{;b} - \frac{1}{2} \, g_{ab}\, 
\phi_{;c}\phi^{;c}\right]
    \nonumber \\
  + \frac{1}{\phi}\; \left[\phi_{;ab}-g_{ab}
{\phi_{;c}}^{;c}\right]+8\pi G \;
\frac{T_{ab}}{\phi} \,, 
\end{eqnarray}
and
\begin{eqnarray}
\Box{\phi}+\frac{{2\phi^2\lambda'(\phi)-2\phi\lambda(\phi)}}
{{2\omega(\phi)+3}}
 = \frac{1}{2\omega(\phi)+3} \times
   \nonumber \\
\times \left[ 8\pi G\, T-\omega'(\phi) 
\phi_{;c}\phi^{;c}
\right] \,,
\end{eqnarray}
where $T\equiv T^{a}{}_{a}$ is the trace of the matter energy-momentum tensor and $G \equiv (2\omega+4)/(2\omega+3)$ is the gravitational constant normalized to the Newton value.
Additional to these equations, one also requires diffeomorphism invariance and consequently the conservation of the matter content $\nabla^b T_{ab}=0$. The latter also preserves the equivalence principle.
Brans-Dicke theory is characterized by the restriction of $\omega(\phi)\,$ 
being a constant, and of $\lambda=\lambda'=0\,$. 

According to the discussion in the previous section, for the general  class of scalar-tensor theories, the tensor term $H_{ab}$ is defined by
\begin{eqnarray}
H_{ab}=-\frac{\omega(\phi)}{\phi^2}\;
\left[\phi_{;a}\phi_{;b} - \frac{1}{2} \, g_{ab}\, 
\phi_{;c}\phi^{;c}\right]
  \nonumber \\
  - \frac{1}{\phi} \left[\phi_{;ab}-g_{ab}
{\phi_{;c}}^{;c}\right]-\lambda(\phi)g_{ab} \,, \label{H-ab_ST}
\end{eqnarray}
and the coupling functions are given by $g_1(\Psi^i)=\phi$, which we shall assume positive, and $g_2(\Psi^i)=1$.  The above considerations on the energy conditions straightforwardly apply. In particular 
Eq. (\ref{en_cond_strong_2}) is easily recovered  like the other energy conditions. 
Taking into account the assumption $\phi>0$, the condition $R_{ab}\,W^a\,W^b\ge 0 $, that yields the focusing of the time-like congruence becomes 
\begin{eqnarray}
( T_{ab}  - \frac{1}{2}\,g_{ab}\,T)\,W^aW^b \ge \phi\,( H_{ab}  - \frac{1}{2}\,g_{ab}\,H)\,W^aW^b \; . \label{en_cond_strong_ST1}
\end{eqnarray}
Notice that even in the presence of a mild violation of the energy condition, the satisfaction of the above condition allows for the focusing of the time-like paths. This is an interesting result since matter may exhibit unusual thermodynamical features, for instance, the presence of negative pressures, and yet gravity retains its attractive character. Alternatively, we see that repulsive gravity may occur for common matter, i.e., for matter that satisfies all the energy conditions. This happens when $H_{ab}$ has the reverse sign in (\ref{en_cond_strong_ST1}).
The energy conditions in the Jordan frame was considered in \cite{Chatterjee:2012zh}, where the null energy condition, in its usual form, can appear to be violated by transformations in the conformal frame of the metric.

The decomposition (\ref{redef_tilderho})--(\ref{redef_tilderho-d}) of the tensor $H_{ab}$ into components parallel to the time-like vector flow $W^a$ and orthogonal to it, is given by the following relationship
\begin{eqnarray}
H^{ab} &=& H_{||} W^aW^b + H_{\bot} h^{ab} + 2\,H_{\bot}^{(a}\, W^{b)} + H_{\bot}^{<ab>}
   \nonumber  \\
&=& \frac{1}{\phi}\,\left[  \tilde\rho W^aW^b + \tilde p h^{ab} + 2\,\tilde q^{(a}\, W^{b)} + \tilde\pi^{ab}\right]
\end{eqnarray} 
where 
$H_{||}$ and $H_{\bot}$ are scalars, $H_{\bot}^{a} $ is a vector and $H_{\bot}^{<ab>}$ is a projected trace-free symmetric tensor (PSTF). This decomposition permits to translate the condition (\ref{en_cond_strong_ST1}) into
\begin{equation}
\frac{1}{\phi}\,(\rho+3p)-(H_{||}+3H_{\bot}) \ge 0 \; .
\end{equation}
In the latter expression we have used
\begin{eqnarray}
H_{||} &= & - \frac{\omega(\phi)}{2\phi^2}\,\left(3\dot\phi^2-h^{cd}\,\nabla_c\phi\, \nabla_c\phi\right) 
   \nonumber \\ 
&& -\frac{1}{\phi}\,h^{cd}\nabla_c\nabla_d\phi +\lambda(\phi) \,,  \\
H_{\bot} &=&- \frac{\omega(\phi)}{3\phi^2}\,\left( \frac{\dot\phi^2}{2}-\frac{1}{2}h^{cd}\,\nabla_c\phi\, \nabla_c\phi\right)  
  \nonumber \\ 
&&
-  \frac{1}{2\phi}\,\left(W^aW^b \, \nabla_c\nabla_d\phi -\frac{1}{3}\,h^{cd}\nabla_c\nabla_d\phi\right) - \lambda(\phi)  \,.
\end{eqnarray}

Clearly, gravity is repulsive or attractive depending on the functions $\omega(\phi)$ and $\lambda(\phi)$. Indeed,  Eq. (\ref{en_cond_strong_2}) reads
\begin{eqnarray}
W^aW^b\,R_{ab}-\frac{\omega(\phi)}{\phi^2}\;
\left(\phi_{;a}\phi_{;b} - \frac{1}{2} \, g_{ab}\, 
\phi_{;c}\phi^{;c} \right)
  \nonumber \\  
- \frac{1}{\phi} \left(\phi_{;ab}-g_{ab}{\phi_{;c}}^{;c}\right)
- \lambda(\phi)g_{ab} =
\nonumber \\  
 W^aW^b\,\frac{8\pi}{\phi}\,\left( T_{ab}-\frac{1}{2}\,g_{ab}\,T\right)  \ge 0
 \; , \label{en_cond_strong_ST0}
\end{eqnarray}
which amounts to
\begin{eqnarray}
 W^aW^b  \Bigg[ \frac{8\pi}{\phi}\,\left(T_{ab}-\frac{\omega+1}{2\omega+3}\,g_{ab}\,T\right) 
+\frac{\omega}{\phi^2}\nabla_a\phi\nabla_b \phi 
   \nonumber \\ 
+\frac{\nabla_a\nabla_b \phi}{\phi}    -\frac{1}{2\phi}\frac{\omega'}{2\omega+3}\,g_{ab}\nabla_c\nabla^c \phi 
  \nonumber  \\
+ g_{ab}\,\frac{\phi\lambda'-(\omega+1)\lambda}{2\omega+3}\Bigg] \ge 0  \; . \label{en_cond_strong_ST4}
\end{eqnarray}

Considering a Friedmann-Lema\^{\i}tre-Robertson-Walker (FLRW) universe, we derive the following inequality 
\begin{equation}
\frac{8\pi G}{\phi}\,\frac{(\omega+3)\rho+3\omega p}{2\omega+3} +\frac{\lambda}{3} + \frac{\omega}{3}\,\frac{\dot\phi^2}{\phi^2}+ \frac{\dot\omega}{2(2\omega+3)}\,\frac{\dot\phi}{\phi} +H\frac{\dot\phi}{\phi}\ge 0 \; .
\end{equation} 
This result shows how the functions $\omega(\phi)$ and $\lambda(\phi)$ define whether gravity is attractive or repulsive in the scalar-tensor cosmological models.

Furthermore, upon a conformal transformation of the theory into the so-called Einstein frame, using $g_{ab}\to \bar g_{ab} = (\phi/\phi_\ast)\, g_{ab}$,  the  condition for gravity to be attractive with the redefined Ricci tensor becomes
\begin{equation}
\tilde R_{ab} u^a u^b= \frac{4\pi}{\phi_\ast} \, (\bar\rho+3\bar p) + \frac{8\pi}{\phi_\ast}\, \left[\dot{\varphi}^2-\tilde{V}(\varphi)\right]\ge 0\;.
\label{ineqSTT}
\end{equation}
Here $\varphi= \int \sqrt{(2\omega+3)/2}\, {\rm d}\ln\phi$ is the redefined scalar field,  $V(\varphi)= \lambda(\phi(\varphi))/\phi(\varphi)$ is the rescaled potential,  $\bar\rho = \rho/\phi^2$, $\bar{p}=p/\phi^2$, and $\phi_\ast$ is an arbitrary value of $\phi$ that guarantees, on the one hand,  that  the  conformal factor is dimensionless,  and, on  the other hand, that it might be related to Newton's gravitational constant $G_N$ by setting $\phi_\ast =G_N^{-1}$. Despite the fact that the inequality (\ref{ineqSTT}) adopts the familiar form found in general relativistic models endowed with a combination of matter and a scalar field, the role of the functions $\omega(\phi)$ and $\lambda(\phi)$ underlies the result because the definitions of $\varphi$ and $V(\varphi)$ depend on them. Another interesting feature, in the Einstein frame, is that the matter and the scalar field are interacting with each other as revealed by the scalar field equation
\begin{equation}
\ddot\varphi+\bar{\theta}\dot\varphi= -\frac{\partial V(\varphi)}{\partial \varphi}- \frac{\partial \bar\rho(\varphi,\bar{a})}{\partial \varphi}\; .
\end{equation}
So  the dependence on the parameters that underlie, on the one, the shape of the self-interacting potential $V(\varphi)$, and  on the other hand, the coupling $\partial_\varphi \bar{\rho}\propto \alpha(\varphi) a^{-3\gamma}$, where $\alpha=(\sqrt{2\omega+3})$, when considering a  perfect fluid with $\bar p=(\gamma-1)\bar\rho$.

In a cosmological setting, gravity may exhibit a transition from being attractive into becoming repulsive when the interplay between the intervening components is such that those which violate the strong energy condition become dominating. The typical case is provided when $V(\varphi)$ has a non-vanishing minimum \cite{ETG_2_GR}.

\subsection{$f(R)$ gravity}

The action in this case is 
\begin{equation}
S=\frac{1}{16\pi} \int \sqrt{-g} f(R) d^4x  + S_M \; ,
\end{equation}
where $R$ is the Ricci scalar (we refer the reader to \cite{Sotiriou:2008rp} for further details).
The $H_ {ab}$ term includes non-linear combinations of the curvature invariants built from the Riemann and Ricci tensors  as well as from derivatives of these tensors, and the couplings $g_1(\Psi^i)= F(R) = f'(R)$ and $g_2(\Psi^i)=1$, where the prime is  the derivative with respect to $R$. In fact, the gravitational field equation is given by
\begin{eqnarray}
F(R) \, G_{ab} + \frac{1}{2}\,\left[R F(R)-f(R)  \right]\, g_ {ab} 
- \nabla_a\nabla_b F(R) 
   \nonumber \\
+g_ {ab}\, \Box F(R) = 8\pi G \, T_ {ab} \,, 
\end{eqnarray}
which can be recast as
\begin{eqnarray}
 G_{ab} = 8\pi G \,\left(\frac{ T_ {ab}}{F(R)}\right)-  \frac{1}{F(R)}\,\Bigg[\frac{1}{2}\,\left(R F(R)-f(R)  \right)\, g_ {ab} 
    \nonumber  \\ 
 - \nabla_a\nabla_b F(R) +g_ {ab}\, \Box F(R) \Big]  \, ,
 \label{FRfieldeq}
\end{eqnarray}
so that we identify 
\begin{eqnarray}
H_ {ab} =\frac{1}{F(R)}\, \Bigg \{\frac{1}{2}\,\left[R F(R)-f(R)  \right]\, g_ {ab} 
- \nabla_a\nabla_b F(R) 
   \nonumber \\
+g_ {ab}\, \Box F(R) \Big\} \,.
\end{eqnarray}
Note that as before ${\nabla}_{a}$ is the covariant derivative operator associated with $g_{ab}$, $\Box \equiv g^{ab} {\nabla}_{a} {\nabla}_{b}$ is the covariant d'Alembertian, and $T^{M}_{ab}$ is the contribution to the stress energy tensor from ordinary matter.
Clearly the above considerations hold completely and gravity is attractive or repulsive depending on the form of $f(R)$.

In the present case we have
\begin{eqnarray}
H_{||} &=& - \frac{1}{F}\,  \left[\frac{1}{2}\,( RF-f) - h^{cd}\,\nabla_c\nabla_d F \right] \,, \\  
H_{\bot} &=&\frac{1}{F}\,\left[ \frac{1}{2}(RF-f)-\frac{1}{3}h^{cd}\,\nabla_c  \nabla_c F +\Box F \right] \, ,
\end{eqnarray}
so that gravity is attractive when
\begin{eqnarray}
8\pi G(\rho+3p) \ge  \left[ ( RF-f) - 2 h^{cd}\nabla_c\nabla_d F + 3\Box F\right] \,.
\end{eqnarray}
Note, however, that this latter condition is still not a condition on any initial data or on matter $T_{\mu\nu}$. Indeed, the higher derivatives may still be eliminated using
the equations of motion; thus it is not an energy condition.

This condition reduces to the usual $(\rho+3p) \ge0$ when $f\propto R$ and hence GR is recovered. More importantly it reveals how the non-linear terms in the action induce attractive or repulsive effects. If there were no matter, i.e., in a vacuum setting gravity becomes repulsive if
\begin{equation}
 ( RF-f) - 2 h^{cd}\nabla_c\nabla_d F + 3\Box F\le 0\;.
\end{equation} 
We refer the reader to \cite{Albareti:2012va, Albareti:2014dxa} for considerations on the non-attractive character of gravity in $f(R)$ theories.

If instead of the strong energy condition we evaluate the null energy condition $R_{ab}k^ak^b\ge 0$, there is once again a considerable simplification of the equations, and we obtain focusing of light bundles when
\begin{eqnarray}
T_{ab} \,k^ak^b + k^ak^b\frac{\nabla_a\nabla_b F}{F(R)} \ge 0 \; .
\end{eqnarray} 
This is a kind of Poisson-like inequality which effectively yields the lensing effect.


We emphasize that in a cosmological setting, the above considerations are particularly important, as in GR the presence of dark energy implies the violation of specific energy conditions. However, in the generalized approach outlined above, there is no violation but just a reinterpretation of the further degrees of freedom emerging from dynamics. 

For instance, consider a flat FRW metric given by $ds^2=-dt^2+a^2(t)\left[dr^2+r^2(d\theta^2+\sin^2\theta \,d\phi^2)\right] $, so that Eq. (\ref{FRfieldeq}) immediately yields the following the field equations
\begin{eqnarray}
\left(\frac{\dot{a}}{a}\right)^2-\frac{1}{3F(R)}\Big\{\frac{1}{2}
\left[f(R)-RF(R)\right]
      \\ \nonumber 
-3\left(\frac{\dot{a}}{a}\right)\dot{R}
F'(R)\Big\} =\frac{\kappa}{3}\rho \,,
\end{eqnarray}
\begin{eqnarray}
\left(\frac{\ddot{a}}{a}\right)+\frac{1}{2F(R)}\Big\{\frac{\dot{a}}{a}\dot{R}
F'(R)+\ddot{R}F'(R)+\dot{R}^2F''(R)
    \nonumber \\
-\frac{1}{3}\left[f(R)-RF(R)\right]\Big\}
=-\frac{\kappa}{6}(\rho+3p) \,.
\end{eqnarray}

Indeed, in the literature, these field equations are usually written as effective Friedman equations, in the following form
\begin{eqnarray}
\left(\frac{\dot{a}}{a}\right)^2&=&\frac{\kappa}{3}\rho_{\rm tot} \,,  \\
\left(\frac{\ddot{a}}{a}\right)&=&-\frac{\kappa}{6}(\rho_{\rm
tot}+3p_{\rm tot}) \,,
    \label{rho+3p}
\end{eqnarray}
where $\rho_{\rm tot}=\rho+\rho_{(c)}$ and $p_{\rm tot}=p+p_{(c)}$, and the quantities
$\rho_{(c)}$ and $p_{(c)}$, are defined as
\begin{eqnarray}
\rho_{(c)}=\frac{1}{\kappa F(R)}\left\{\frac{1}{2}
\left[f(R)-RF(R)\right]-3\left(\frac{\dot{a}}{a}\right)\dot{R}
F'(R)\right\} \,, \nonumber \\
p_{(c)}=\frac{1}{\kappa
F(R)}\Bigg\{2\left(\frac{\dot{a}}{a}\right)\dot{R}
F'(R)+\ddot{R}F'(R)+\dot{R}^2F''(R)
    \nonumber  \\
-\frac{1}{2}\left[f(R)-RF(R)\right]\Big\}
 \,, \nonumber
\end{eqnarray}
respectively. However, one should always bear in mind that these quantities have a geometrical origin, and should not be interpreted as a fluid.

Now, from Eq. (\ref{rho+3p}), it is transparent that an accelerated expansion can be obtained by imposing the condition $\rho_{\rm tot}+3p_{\rm tot}<0$. Note that, in principle, one may impose that normal matter obeys all of the energy conditions, and the acceleration $\ddot{a} \geq 0$ is attained by considering an appropriate functional form for $f(R)$. For simplicity, consider vacuum, $\rho=p=0$, so that the energy conditions are border-line satisfied. Now appropriately defining a parameter $\omega_{\rm eff}=p_{(c)}/\rho_{(c)}$, one may impose a function $f(R)$. For instance, consider the model $f(R)=R-\mu^{2(n+1)}/R^n$ analysed in \cite{Carroll:2003wy}. By choosing a generic power law for the scale factor, the parameter can be written as
\begin{equation}
\omega_{\rm eff}=-1+\frac{2(n+2)}{3(2n+1)(n+1)} \,,
\end{equation}
and the desired value of $\omega_{\rm eff}<-1/3$ may be attained, by appropriately choosing the value of the parameter $n$.

We emphasize that the message that one obtains from this analysis is precisely that in the generalized approach outlined in this work, there are no violation of the GR energy conditions, but just a reinterpretation of the further degrees of freedom emerging from the dynamics.

\section{Summary and Discussion}\label{sec:concl}

In this work, we have considered the further degrees of freedom related to curvature invariants and scalar fields in Extended Theories of Gravity (ETG). These new degrees of freedom can be recast as {\it effective fluids} that carry different meanings with respect to the standard matter fluids generally adopted as sources of the field equations. It is thus somewhat misleading to apply the standard general relativistic energy conditions to this effective energy-momentum, as the latter contains the matter content and  geometrical quantities, which arise from the particular ETG considered. It can be shown, as done in Sec. \ref{secIII}, that the further dynamical content of ETG can be summed up  into two coupling functions $g_{1}$ and $g_{2}$ and an additional tensor $H_{ab}$ where all the geometrical modifications are present. Clearly GR is immediately recovered as soon as $g_1=g_2=1$ and $H_{ab}=0$.  Here, we explored these features to cases with the contracted Bianchi identities with diffeomorphism invariance and to cases with generalized explicit curvature-matter couplings, which imply the non-conservation of the energy-momentum tensor. Furthermore, we applied the analysis to specific ETGs, such as scalar-tensor gravity and $f(R)$ gravity. The main outcomes are that  matter can  exhibit further thermodynamical features and gravity can retain its attractive character in  presence of large negative pressures.  On the other hand, repulsive gravity may occur for standard matter. 

As a general result, the fact that further degrees of freedom, related to ETG, can be dealt under the standard of effective 
fluids allows, in principle,  to set consistent energy conditions for large classes of theories. In this sense, the well formulation of the Cauchy problem can be considered a standard feature for several theories of gravity. 
From a cosmological point of view, these considerations are crucial. For example, the presence of dark energy can be considered a straightforward violation of energy conditions in the standard  sense of GR. In our generalized approach, there is no violation but just a reinterpretation of the further degrees of freedom emerging from dynamics.

  
Furthermore, a few considerations in the context of scalar-tensor theories are in order. Note that in the Einstein frame one verifies that the energy conditions are satisfied, but may be violated in the Jordan frame \cite{magnano,Deser:1983rq,Faraoni:2004pi}. This fact does not eliminate the presence of singularities when both frames are considered equivalent (see below for a discussion on the latter issue) \cite{Tsamparlis:2013aza}. Thus, in order to avoid these ambiguities, one may wonder that due to the fact that the energy conditions essentially hold in relativity, why not restrict oneself to the Einstein frame formulation (31) and not bother about the geometrical or matter nature of the appropriate quantities?
However, it is important to mention that in some specific situations it is also possible that the weak energy condition is satisfied in the Jordan frame \cite{Faraoni:2004pi}, and thus evades the problems mentioned above. In addition to this, there are situations, where it is useful to work in the Jordan frame. For instance, if one uses the Equivalence Principle (EP) as a guide in constructing one's theory, then it is useful to work in the Jordan frame, as here the EP is satisfied, and the latter is violated in the Einstein due to the fifth force arising as a result of the anomalous coupling of the scalar field to matter. Nevertheless, one may argue that this may be misleading as the EP could indeed be violated in nature, provided that the violations are extremely small in order to evade detection from current measurements, and thus serve to place stringent constraints on theories that imply the non-conservation of the energy-momentum tensor, and that consequently manifest non-geodesic motion. One may also mention that if one only restricts attention to the Einstein frame, one may also lose sight of the original motivations and modifications of gravity in the geometrical sector. Indeed, the conformal transformation mixes the geometric and matters degrees of freedom, which results in many interpretational ambiguities \cite{Capozziello:1996xg}. 
Furthermore, note that Dicke's argument is purely classical, and in this respect, at the quantum level the equivalence of both frames is not proven. In fact, when the metric is quantized, one can find inequivalent quantum theories \cite{Grumiller:2003mc}. In addition to this, considering the semi-classical regime, in which gravity is classical and the matter fields are quantized, one would also expect that the conformal frames are inequivalent, and we refer the reader to \cite{Tsamparlis:2013aza} (and references therein) for more details.

The viewpoint that the Einstein and Jordan frames are physically equivalent is correct and consistent, and can be traced back to Dicke's original paper \cite{Dicke:1961gz}, where the conformal transformation technique was introduced. Indeed, in the spirit of Dicke's paper, both conformal frames are equivalent provided that in the Einstein frame the units of mass, time and space scale as appropriate powers of the scalar field, and are thus varying. More specifically, physics must be conformally invariant and the symmetry group of gravity should be enlarged to incorporate conformal conformations, in addition to the group of diffeomorphisms \cite{Faraoni:2004pi}. However, it is common practise in the literature to consider that in the Einstein frame, measurements are referred to in a rigid system of units, instead of units varying with the conformal factor, and consequently resulting in the non-equivalence of the Jordan and the Einstein frames \cite{Faraoni:2004pi}. Although this approach is perfectly legitimate from a mathematical point of view, one should keep in mind that both theories are physically inequivalent, for instance, when one considers cosmological or black hole solutions. The issue then becomes which of the two conformal frames is physical? In the context of the energy conditions, these are satisfied in the Einstein frame, and violated in the Jordan frame. 
For instance, in this context, the violations of the weak energy condition in the Jordan frame is also responsible for the violation of the second law of black hole thermodynamics \cite{Faraoni:2004pi} (and  references therein). In fact, if the weak energy condition is violated, the Hawking-Penrose singularity theorems \cite{Hawking:1973uf} also do not apply in the original Jordan frame. 
In order to circumvent this difficulty, one may consider the approach outlined in \cite{Chatterjee:2012zh}, in that the second law of black hole thermodynamics is taken as fundamental, and then one modifies the null energy condition in a given theory of gravity to ensure that the classical black hole solution has an entropy that increases with time. This approach seems appealing as the null energy condition does not seem to rest on any fundamental principle of physics, unlike the second law of black hole thermodynamics.

\section*{Acknowledgements}

SC acknowledges the INFN (iniziative specifiche TEONGRAV and QGSKY). FSNL is supported by a Funda\c{c}\~{a}o para a Ci\^{e}ncia e Tecnologia Investigador FCT Research contract, with reference IF/00859/2012, funded by FCT/MCTES (Portugal). FSNL and JPM acknowledge financial support of the Funda\c{c}\~{a}o para a Ci\^{e}ncia e Tecnologia through the grants CERN/FP/123618/2011 and EXPL/FIS-AST/1608/2013.


\end{document}